\documentclass[amsmath,amssymb,aps,prd,nofootinbib]{revtex4}
\usepackage[utf8]{inputenc}
\pdfoutput=1 
\usepackage{graphics,graphicx}
\usepackage[usenames,dvipsnames]{color}

\def\bma{\left( \begin{array} }
\def\ema{\end{array} \right)}

\begin{document}

\preprint{\hfill{KIAS-P13054}}

\title{Axion Dark Matter with High-Scale Inflation}
\author{Eung Jin Chun}
\affiliation{
Korea Institute for Advanced Study, Seoul 130-722, Korea}

\begin{abstract}
We show that supersymmetric axion models breaking the PQ symmetry by the interplay of non-renormalizable supersymmetric 
terms and soft supersymmetry breaking terms  provide a natural framework not only for generating the axion scale from soft supersymmetry breaking scale $m_{3/2}$ but also for enhancing it during inflation by factor of order $\sqrt{H_I/m_{3/2}}$ where $H_I \simeq 10^{14}$ GeV according to the recent BICEP2 result.
In this scheme, the PQ symmetry can stay broken throughout the whole history of the Universe 
if the reheat temperature is below $10^{10}$ GeV, or $m_{3/2}$ when the PQ fields couple strongly to
thermal (Standard Model) particles. It is also shown that parametric resonance during preheating is  
not effective enough to induce non-thermal PQ symmetry restoration.
As a consequence, axion models with the QCD anomaly $N_{DW}>1$ can be made free from
the domain wall problem while the axion isocurvature perturbation is suppressed sufficiently 
for the axion scale during inflation larger than about 
$M_P (\Omega_a h^2/0.12)^{1/2} ( F_a/10^{12} \mbox{GeV})^{0.6}$ GeV.
\end{abstract}

\maketitle


{\it Introduction:}
The recent measurement of CMB B-mode polarizations by BICEP2 points to 
large primordial tensor perturbations with the tensor-to-scalar ratio 
$r ={\cal O}(0.1)$  \cite{bicep2}. This reveals a high-scale inflation with 
the Hubble parameter,
\begin{equation} \label{HI}
 H_I \simeq 10^{14} \mbox{ GeV} \left( r \over 0.16\right)^{1/2}
\end{equation}
which corresponds to the inflation energy scale of $V_I^{1/4}\equiv (3 H_I^2 M_P^2)^{1/4}
 \simeq 2.7\times 10^{16} (r/0.16)^{1/4}$ GeV. 
Furthermore, the chaotic inflation with a quadratic potential \cite{linde83} 
seems to fit nicely other CMB observables measured by Planck \cite{planck} 
If confirmed, it has profound implications for the axion dark matter \cite{higaki,marsh,visinelli,dias}.
In this work, we present a natural framework where the PQ symmetry is never restored throughout 
the whole history of the Universe, and the PQ symmetry breaking scale during inflation 
is much larger than the present one so that the domain wall problem occurring in axion models with
the QCD anomaly $N_{DW}>1$ can be avoided, and the axion isocurvature density perturbation can be sufficiently 
suppressed.

\bigskip

{\it Axion and strong CP problem:}
The axion was introduced to resolve the strong CP problem in a dynamical way (for a recent review, see, \cite{kim08,kawasaki}). Being a pseudo-Goldstone boson of a QCD-anomalous $U(1)$ symmetry (PQ symmetry)
the axion couples to the CP-odd QCD field strength term;
\begin{equation} \label{aGG}
 {\cal L}_{QCD} = {a \over  F_a} {g_s^2 \over 32 \pi^2}  G^a_{\mu\nu} \tilde{G}_a^{\mu\nu}
\end{equation}
where $F_a$ is the axion decay constant and  
$\tilde G_a^{\mu\nu} = {1\over2} \varepsilon^{\mu\nu\rho\sigma} G^a_{\rho\sigma}$.
That is, the QCD $\theta$ angle is replaced by a dynamical field $a$: $\theta \equiv a/F_a$. 
After the QCD phase transition, instanton effect generates the axion potential,
\begin{equation} \label{Va}
 V(a) = m_a^2 F_a^2 \left[ 1- \cos{a\over F_a}\right]
\end{equation}
where the axion mass is given by $m_a = \sqrt{Z}/(1+Z) (f_\pi m_\pi / F_a)$ with the the up 
and down quark mass ratio $Z=m_u/m_d$, and $m_\pi$ ($f_\pi$) is the pion mass (decay constant). The axion potential 
sets the vacuum expectation value $\langle a \rangle =0$ ensuring no $\theta$ contribution to 
the neutron electric dipole moment.  

\bigskip

{\it Axion cold dark matter:}
The occurrence of the axion potential (\ref{Va}) also plays an important role in cosmic axion production.
If the PQ symmetry is broken after inflation, there appear three sources of axion production: coherent oscillation 
from initial misalignment by $\theta_i$, axionic string formation upon the PQ symmetry breaking,
and domain wall production during the QCD phase transition.
Summing these up one gets \cite{kawasaki,visinelli}
\begin{equation} \label{Oa}
\Omega_a h^2 \approx 0.18\, \langle \theta_i^2 \rangle\, \alpha_{\rm t.d.} 
\left( F_a \over 10^{12} \mbox{ GeV}\right)^{1.19}
\left( \Lambda \over 400 \mbox{ MeV} \right)
\end{equation}
where $\langle \theta_i^2 \rangle \approx 2 \pi^2/3$ is a value averaged over possible ranges of  
the initial misalignment angle $\theta_i$ which is not uniform in our Hubble volume, and 
$\alpha_{t.d.}$ takes into account contributions from strings and domain walls. 
One then finds that the axion cold dark matter (CDM) satisfies
$\Omega_a h^2 \simeq 0.12$ for $F_a \approx 1.4\times 10^{11}\, \alpha_{t.d.}^{-0.84}$ GeV
(see \cite{visinelli} and references therein for different estimates of $\alpha_{t.d.}$). 

Note that the $\theta$ angle defined in (\ref{aGG}) by $\theta \equiv a/F_a$  has a periodicity
$\theta \equiv \theta + 2 \pi N_{DW}$ where $N_{DW}$ is the QCD anomaly, and thus there appear 
distinct $N_{DW}$ $\theta$ vacua: $0, 2\pi, \cdots 2\pi(N_{DW}-1)$ in the axion potential (\ref{Va}).
It is then required to have $N_{DW}=1$ in order not to produce stable string-domain wall networks which
overcloses the Universe.
The KSVZ model with one pair of heavy quarks has $N_{DW}=1$ while
the original DFSZ model has $N_{DW}=6$ \cite{kim08}. 
Let us remark that various hybrid models can be constructed to obtain $N_{DW}=1$.

\bigskip

If the PQ symmetry is broken before or during inflation, axionic strings formed during the PQ phase transition
are efficiently diluted away and no domain wall can form. 
Thus, any axion model with $N_{DW}>1$ is allowed, and the axion dark matter can be produced from 
the coherent oscillation as well as from (massless) axion fluctuations
during inflation, $\delta a \approx H_I/2\pi$. This leads to
\begin{equation} \label{OaI}
\Omega_a h^2 \approx 0.18 \left[ \theta_i^2 + \left( H_I \over 2\pi F_{I} \right)^2 \right]
\left( F_a \over 10^{12} \mbox{ GeV}\right)^{1.19}
\left( \Lambda \over 400 \mbox{ MeV} \right) \,.
\end{equation}
Here a uniform initial misalignment angle $\theta_i$ is taken
as our Universe might be expanded from a small patch with given $\theta_i$ during inflation,  
and the axion decay constant during inflation $F_I$ is assumed to be different from
the present value $F_a$ and thus could suppress the fluctuation contribution to evade the problem 
with axion isocurvature perturbation \cite{linde90,linde91}.

The axion fluctuations produced during inflation become isocurvature density perturbations of the axion 
dark matter after the axion acquires mass at the QCD phase transition.
The recent Planck measurements constrain the isocurvature power spectrum ${\cal P}_a$ 
of the axion CDM compared to the scalar power spectrum ${\cal P_R}$ \cite{planck}:
\begin{equation} \label{Pa}
{\cal P}_a \equiv  4 \xi^2 { (H_I/2\pi)^2 \over ( F_I \theta_i)^2 + (H_I/2\pi)^2 } \lesssim 0.04\, {\cal P_R}
\end{equation}
where $\xi \equiv \Omega_a/\Omega_{CDM}$, and  ${\cal P_R} \simeq 2.2 \times 10^{-9}$.
This essentially rules out the axion as a CDM candidate \cite{marsh,visinelli} if $F_I = F_a$.
In order to see how much the severe constraint from the isocurvature perturbations can be relaxed for $F_I \gg F_a$, 
let us first consider two regions separately: (i) $F_I \theta_i < H_I/2\pi$, and (ii) $F_I \theta_i > H_I/2\pi$.
In the case (i) having ${\cal P}_a \approx 4 \xi^2$, Eqs.~(\ref{OaI}) and (\ref{Pa}) require $\xi < 4.7 \times 10^{-6}$
together with
\begin{eqnarray} \label{lowF}
&& F_I \gtrsim 9 \times 10^{15} \left( F_{a} \over 10^{12} \mbox{GeV} \right)^{0.6}\!\mbox{GeV}, \\
&\mbox{and}\;\;& \theta_i \lesssim 0.0018 \left( 10^{12}\mbox{GeV} \over F_a \right)^{0.6}. \nonumber
\end{eqnarray}
In the case (ii), one finds the requirements:
\begin{eqnarray} \label{highF}
&&F_I \gtrsim 4.2 \times 10^{18}\sqrt{\xi} \left(F_a \over 10^{12}\mbox{GeV}\right)^{0.6}\! \mbox{GeV}, \\
&\mbox{and}\;\;&\theta_i \approx 0.82 \sqrt{\xi} \left(10^{12}\mbox{GeV} \over F_a \right)^{0.6}. \nonumber
\end{eqnarray}
Thus, the axion CDM with $\xi=1$ can be obtained for $F_a \sim 10^{12}$ GeV and $\theta_i \sim 1$ 
if $F_I \sim M_P$ is allowed.

\bigskip

{\it PQ symmetry and supersymmetry breaking:}
The axion decay constant changing with time is a generic feature as far as there is no particular reason to forbid  
inflaton coupling to the PQ fields. A simple and systematic framework can be obtained
in the context of supersymmetry which has to be broken by the inflaton potential of order $V_I$ 
during inflation \cite{dine95}.
Furthermore, the PQ symmetry breaking scale can be correlated to the supersymmetry breaking scale
in a flat potential model with the superpotential \cite{murayama92,choi96}:
\begin{equation} \label{WPQ}
W_{PQ} = \lambda {P^{n+2} Q \over M_P^n}
\end{equation}
where $n$ is a positive integer, and the $U(1)_{PQ}$ charge 1 and $-(n+2)$ are assigned to $P$ and 
$Q$, respectively.  Recall that the dynamical generation of the axion scale from soft supersymmetry breaking terms
can also resolve the $\mu$ problem  by introducing a Higgs bilinear operator
$W_{KN} = h P^{n+1} H_u H_d /M_P^n$  \cite{kim84} which ensures $\mu \sim m_{3/2}$ and realizes
a supersymmetric version of the DFSZ axion model having $N_{DW}=6$.
The potential domain wall problem is evaded
if the PQ symmetry is broken before/during inflation and remains broken 
during the whole history of the Universe as will be discussed below.

In order to deliver the essential features, let us consider a simplified PQ superpotential
together with the quadratic potential for chaotic inflation\footnote{Note that a special form of 
the Kahler potential is required to guarantee the quadratic term in the scalar potential:
$V_I =m^2 |\chi|^2$ \cite{ellis14}.}:
\begin{equation} \label{WIPQ}
W = m_\chi \chi^2 + {\lambda \over n+3} {\phi^{n+3} \over M_P^n}
\end{equation}
where $\chi$ is the inflaton superfield and $\phi$ is a representative of the PQ superfields.
As is well-known \cite{dine95}, a non-renormalizable coupling allowed in the Kahler potential 
\begin{equation}
\delta K \sim {1\over M_P^2} \chi^\dagger \chi \phi^\dagger\phi
\end{equation}
delivers the supersymmetry breaking effect by the inflaton to the PQ sector through F-terms leading 
to the scalar potential $V = V_{soft} + V_{SUSY}$:
\begin{eqnarray} \label{VIPQ}
&& V_{soft} = (C_m H^2 + m_\phi^2) |\phi|^2 + \left[ {(C_A H + A) \lambda \over n+3} 
{ \phi^{n+3} \over M_P^n} + h.c \right], \nonumber \\
&& V_{SUSY} =  |\lambda|^2 {|\phi|^{2n+4} \over M_P^{2n} }
\end{eqnarray}
where $C_m$ and $C_A$ are constants of order one, and $m_\phi$ and  $A$ are the usual soft terms of order $m_{3/2}$. Minimization of the above scalar potential gives rise to the vacuum expectation value
$\langle \phi \rangle= \phi_0$ as follows \cite{chun04}:
\begin{equation} \label{phi0}
\phi_0 \sim \begin{cases}
\left( H_I M_P^n\right)^{1\over n+1} \sim F_I  & \mbox{during inflation} \cr
\left( \mbox{Min}[H, m_\phi] M_P^n \right)^{1\over n+1}  & \mbox{after inflation} \cr
\left( m_\phi M_P^n \right)^{1\over n+1} \sim F_a   & \mbox{after reheating}\,. \cr 
\end{cases} 
\end{equation}  
One can see that the PQ symmetry remains broken during the whole history of the Universe if 
the reheating occurs at the Hubble parameter smaller than $m_\phi$,
which sets an upper limit on the reheat temperature:
\begin{equation}
T_R \lesssim 10^{10} \left( m_\phi \over 200 \mbox{GeV} \right)^{1/2} \mbox{GeV} .
\end{equation}
to maintain the initial PQ symmetry breaking vacuum stays connected to the present one.
The reheat temperature needs to be even lower  if a PQ field has a sizable coupling to
any fields in thermal equilibrium and thus obtains a thermal mass.  For instance, the PQ field may couple to a
right-handed neutrino, $W \sim y \phi NN$ to realize the seesaw mechanism \cite{murayama92,choi96},
or to a heavy quark, $W \sim  y P Q Q^c$ in the KSVZ model. 
These couplings generate thermal mass-squared of order $\delta m^2_\phi \sim y^2 T^2_{R}$ which restores the PQ symmetry 
if $\delta m^2_\phi > m_{\phi}^2$. To avoid this, we put a rough condition of 
\begin{equation}
T_R \lesssim  {m_{\phi} \over y} .
\end{equation}

Another important issue concerning symmetry non-restoration is 
parametric resonance which can leads to huge production of bosonic fluctuations in the process of preheating
during the inflaton oscillation period,
and thus non-thermal symmetry restoration \cite{kofman94,shtanov94}. 
If this happens, subsequent symmetry breaking brings back the topological defect production
and thus the results in (\ref{Oa}) should be applied for $N_{DW}=1$ 
while ruling out axion models with $N_{DW}>1$.
This is a generic feature if the PQ field has a direct coupling to the inflaton: 
$\delta {\cal L} \sim g^2 |\chi|^2 |\phi|^2$.
Even without such a coupling, the PQ symmetry can be restored if there is a large initial value $\phi_i$
during the inflation oscillation period before the reheating, and PQ symmetry non-restoration is ensured if
$F_a \gtrsim 10^{-4} \phi_i$  \cite{kawasaki13}. 
In supersymmetric theories, there appear couplings between the inflaton and PQ fields through the supersymmetry
breaking effect (\ref{VIPQ}) although direct couplings in the superpotential are forbidden.
In this case, parametric resonance is never effective as we will see in the following.
To simplify the analysis, let us take $n=2$ and ignore the A terms ($m_\phi \gg A$ and $C_m \gg C_A$), 
and assume negative mass-squared terms\footnote{For a detailed analysis with general soft terms for the
original superpotential (\ref{WPQ}), see Ref.~\cite{chun00}.}:
\begin{equation}
V= -( C_m H^2 + m_\phi^2) |\phi|^2 + {\lambda^2 \over M_P^2} |\phi|^6.
\end{equation}
When the mass term is constant (during inflation and $H\ll m_\phi$), the homogeneous field value $\phi_0 = \langle \phi \rangle$ is set by
$\phi_0 = \sqrt{ C_m H_I M_P / \sqrt{3} \lambda }$ during inflation, and $\phi_0 = \sqrt{ m_\phi M_P / \sqrt{3} \lambda }$ after reheating for $H \ll m_\phi$.
During the period of inflaton oscillation before reheating
$\phi_0$ follows the equation of motion 
\begin{equation} \label{phi0eq}
\ddot \phi_0 + 3H \dot \phi_0 + {3 \lambda^2 \over M_P^2} \phi^5_0 - C_m H^2  \phi_0=0
\end{equation}
ignoring the $m^2_\phi$ term. It is solved by
\begin{equation} \label{phi0sol}
\phi_0 = A \sqrt{ {H M_P\over \sqrt{3} \lambda} } 
\end{equation}
where $H=2/3t$ and  the coefficient $A$ is determined to be $A = \left( {4\over9} C_m +{1\over4} \right)^{1/4}$
which is slightly different from the ones during inflation and after reheating. 
Treating properly the transition between the inflation (reheating) and oscillation periods, one would be able to find
a smooth transition of $A(t)$ connecting the different values during inflation, inflaton oscillation, and radiation-dominated periods.  For our analysis, we will take a simple approximation of an abrupt change of periods 
and use the solution (\ref{phi0sol}) to find the evolution of fluctuations
of the PQ field: $\phi = \phi_0 + \delta \phi$, during the oscillation period.
The linearized equation of motion for the fluctuations in Fourier space is
\begin{equation} \label{delphi}
 \ddot \delta \phi_k + 3 H \dot \delta \phi_k - \left( {k^2 \over a^2} + \tilde C_m H^2 \right) \delta \phi_k =0 
\end{equation}
where $\tilde C_m = {11 \over 9} C_m + {5\over 4} $. Note that the source term does not contain the oscillating term of the inflaton
contrary to the usual parametric resonance case with a direct coupling between the inflation and PQ fields 
in non-suersymmetric models \cite{kofman94,shtanov94,kawasaki13}. 
Therefore, no parametric resonance can occur and it should remain true even after considering
the transition effect properly.  One can indeed find that
an asymptotic behavior of solutions to the equation (\ref{delphi}) is given by
\begin{equation}
 \delta \phi_k \sim { \sin(k t^{3/2} )  \over k^{4/5} t^{4/3} }
\end{equation}
which is a oscillating and decaying function showing no resonance effect.

\bigskip

Let us finally see how large $F_a \gtrsim 10^{12}$ GeV with $F_I \gtrsim M_P$ in (\ref{highF}) can be allowed 
in our framework. As we have $F_I / F_a= \sqrt{ C_m H_I / m_\phi}$ independently of $n$, 
the first relation of (\ref{highF}) translates to 
\begin{equation}
 C_m \gtrsim 3 \,\xi \left( m_\phi \over 10^2 \mbox{Gev} \right) 
 \left( 10^{13} \mbox{GeV} \over F_a \right)^{0.8} 
\end{equation}
and $F_a \sim 10^{13}$ GeV requires a samll Yukawa coupling of 
$\lambda \sim 10^{-6}$ for $n=2$. Note that one may have $F_a \gg 10^{12}$ GeV allowing trans-Planckian PQ field values during inflation, $F_I \gg M_P$, which, however, does not lead to a dangerous effect in the potential 
as it comes with a small Yukawa $\lambda$.

\bigskip

{\it Conclusion:}
The QCD axion is a well-motivated hypothetical particle which is a pseudo-Goldstone boson of the PQ symmetry 
solving the strong CP problem and is a good candidate for the cold dark matter. 
In generic axion models with the QCD anomaly $N_{DW}>1$, the PQ symmetry needs to be broken before/during
inflation and never restored after inflation in order to prohibit production of stable domain walls.
However, the isocurvature perturbation produced during inflation rules out axions 
as a dark matter candidate if the PQ symmetry breaking scale $F_I$ during inflation is the same as the present one
$F_a$. 
In general, these two symmetry breaking scales need not to coincide, for instance, if 
couplings between inflation and PQ fields are allowed. Supersymmetry provides a natural way
to realize such a situation. Generic order-one couplings between inflaton  and PQ fields allowed 
in the Kahler potential lead to mediation of the inflaton supersymmetry breaking effect to the PQ sector
through F-terms.  
A smooth connection of such an early supersymmetry breaking to the present one can be realized 
if the PQ symmetry breaking is induced by soft supersymmetry breaking terms as presented in this work.
As a consequence, the stringent constraint from isocurvature perturvations can be avoided if $F_I \gtrsim
M_P (\Omega_a h^2/0.12)^{1/2} (F_a/10^{12}\mbox{GeV})^{0.6}$. 

For this to happen, the reheat temperature has to be low enough; $T_R \lesssim 10^{10}$ GeV 
in order to keep the Hubble parameter below the soft supersymmetry breaking scale $m_{3/2}$. 
If PQ fields couple strongly to thermal particles through a coupling $y$,  
it is required to have $T_R \lesssim m_{3/2}/y$ 
in order to forbid the induced thermal masses which can restore the PQ symmetry.
There could also be non-thermal symmetry restoration due to parametric resonance during the inflaton 
oscillation period.  Having the inflaton coupling to the PQ fields only through the Hubble parameter, 
the inflaton supersymmetry breaking effect,
parametric resonance is shown to be ineffective in our framework.

\bigskip
{\it Note added:} The author thanks Kiwoon Choi for sharing their results 
\cite{choi14} which has some overlap consistent with ours.

\end{document}